\documentclass[%
 aip,
 amsmath,amssymb,
 reprint,%
]{revtex4-1}

\usepackage{graphicx}
\usepackage{dcolumn}
\usepackage{bm}

\usepackage[utf8]{inputenc}
\usepackage[T1]{fontenc}
\usepackage{mathptmx}
\usepackage{etoolbox}

\newcommand{\wavenumber}[1]{\ensuremath{#1~\text{cm}^{-1}}}

\makeatletter
\def\@email#1#2{%
 \endgroup
 \patchcmd{\titleblock@produce}
  {\frontmatter@RRAPformat}
  {\frontmatter@RRAPformat{\produce@RRAP{*#1\href{mailto:#2}{#2}}}\frontmatter@RRAPformat}
  {}{}
}%
\makeatother
\begin{document}

\preprint{AIP/123-QED}

\title{Second Harmonic Generation from Grating-Coupled Hybrid Plasmon-Phonon Polaritons}
\author{Marcel Kohlmann}
\affiliation{Fritz-Haber-Institut der Max-Planck-Gesellschaft, Faradayweg 4-6,14195 Berlin, Germany}
\affiliation{Institut für Physik, Universität Greifswald, Felix-Hausdorff-Str. 6, 17489 Greifswald, Germany} 
\author{Christian Denker}
\affiliation{Institut für Physik, Universität Greifswald, Felix-Hausdorff-Str. 6, 17489 Greifswald, Germany} 
\author{Nikolai C. Passler}
 \affiliation{Fritz-Haber-Institut der Max-Planck-Gesellschaft, Faradayweg 4-6,14195 Berlin, Germany}
\author{Jana Kredl}
\affiliation{Institut für Physik, Universität Greifswald, Felix-Hausdorff-Str. 6, 17489 Greifswald, Germany} 
\author{Martin Wolf}
 \affiliation{Fritz-Haber-Institut der Max-Planck-Gesellschaft, Faradayweg 4-6,14195 Berlin, Germany}
\author{Markus M\"unzenberg}
\affiliation{Institut für Physik, Universität Greifswald, Felix-Hausdorff-Str. 6, 17489 Greifswald, Germany} 
\author{Alexander Paarmann}
 \affiliation{Fritz-Haber-Institut der Max-Planck-Gesellschaft, Faradayweg 4-6,14195 Berlin, Germany}
 \email{alexander.paarmann@fhi.mpg.de}

\date{\today}

\begin{abstract}
Polaritons can provide strong optical field enhancement allowing to boost light-matter interaction. Here, we experimentally observe enhancement of mid-infrared second-harmonic generation (SHG) using grating-coupled surface phonon polaritons of the 6H-SiC surface. In our experiment, we measure the SHG along the polariton dispersion by changing the incidence angle of the excitation beam. We observe hybridization between the propagating surface phonon polaritons and localized plasmon resonances in the gold grating, evidenced by the modification of the polariton dispersion as we change the area ratio of grating and substrate. Design options for engineering the plasmon-phonon polariton hybridization are discussed. Overall, we find a rather low yield of polariton-enhanced SHG in this geometry compared to prism-coupling and nanostructures, and discuss possible origins. 
\end{abstract}

\maketitle

The sub-diffractional light localization and enhancement of the local optical fields provided by surface phonon polaritons (SPhPs) \cite{Caldwell2015} make them an ideal platform for applications relying on increased light-matter interaction in the mid- to far-infrared (IR). Typical examples include improved molecular sensing\cite{Folland2020,Zheng2017a,Autore2017,Berte2018}, coherent thermal emission\cite{Greffet2002, Wang2017b,Lu2021}, or nonlinear-optical signal generation\cite{Razdolski2016,Passler2017,Passler2019, Kiessling2019}. Due to their evanescent nature, exciting and probing SPhPs requires specific experimental schemes\cite{Folland2019}, such as nanotip-based near-field mapping\cite{Dai2018,Ma2021}, sub-diffractional nanostructures\cite{Caldwell2013a,Caldwell2014}, prism coupling\cite{Neuner2009,Passler2017,Passler2022}, or grating coupling\cite{Greffet2002,Zheng2017a}. The latter offers several advantages such as extrinsic resonance tuning via the incidence angle\cite{Greffet2002}, far-field access allowing for easy device integration\cite{Zheng2017a,Folland2020} with additional design options using the grating shape\cite{Greffet2002,Folland2020} and material combination of polaritonic substrate and line grating\cite{Zheng2017a}.  

In this work, we demonstrate mid-infrared nonlinear-optical second harmonic generation (SHG) from SPhPs excited via gold-on-6H-SiC line gratings. We observe enhancement of the SHG response at the polariton resonance which can be spectrally tuned through the Reststrahlen band of 6H-SiC by changing the incidence angle. We discuss possible origins for the rather low SHG yield observed. By investigating two different samples with varied Au-to-SiC area ratio, we further observe a modified polariton dispersion which we ascribe to hybridization of the propagating SPhP modes with localized plasmon resonances at the edges of the gold bars. Finite element simulations show excellent agreement with our measurements, and we provide a simple effective medium model qualitatively explaining the plasmon-phonon polariton hybridization. 

\begin{figure*}[bth!]
\includegraphics[width=1\linewidth]{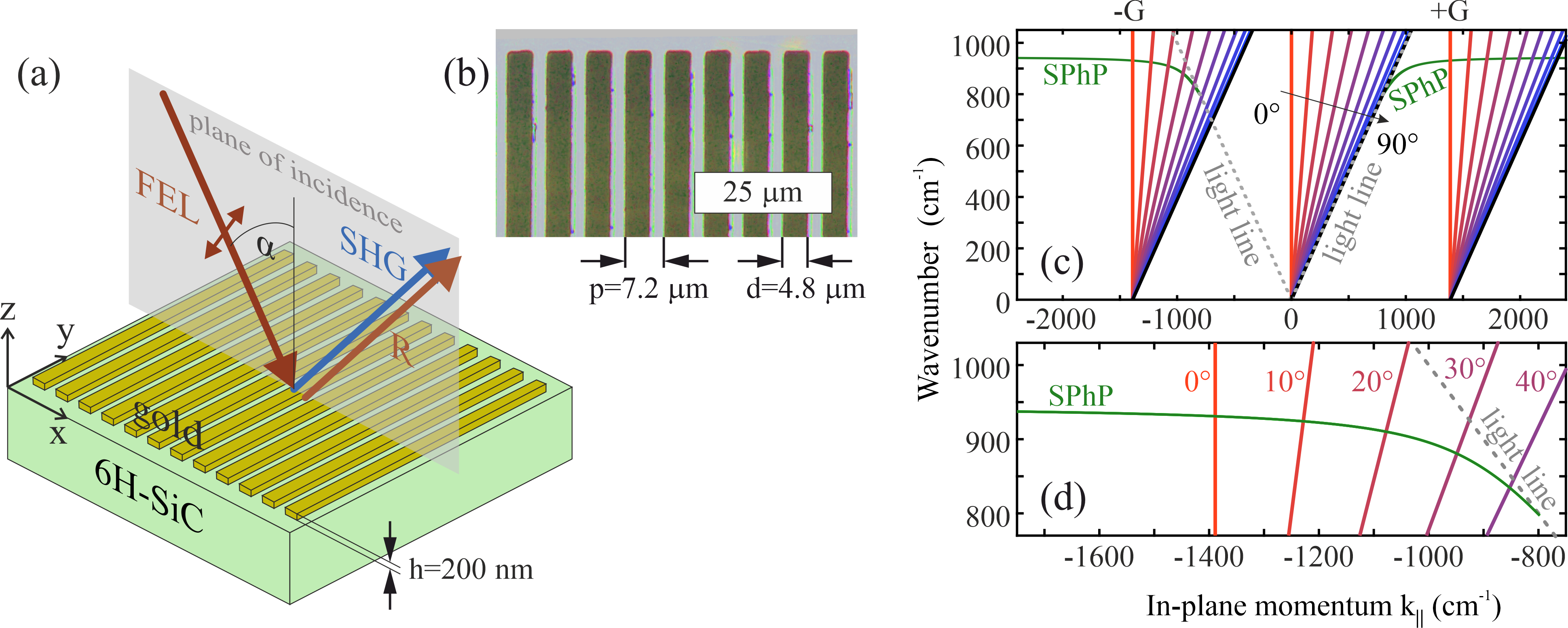}
  \caption{\textbf{Experimental concept for SHG from grating coupled surface polaritons} a) Experimental geometry: wavelength-tunable IR laser pulses polarized in the $x-z$ plane impinge on the 6H-SiC sample with gold grating lines along the $y$-direction, allowing excitation of surface polaritons propagating along the $x$-direction. The specularely reflected IR and SHG intensity are detected simultaneously. A dip in reflectance and a peak in SHG indicates resonant surface polariton excitation. b) Optical microscopy image of the $d$=$4.8$~$\mu$m sample with a period $p=7.2$~$\mu$m. c) Grating coupling to SPhPs: (red-to-blue) angle-of-incidence dispersion of the in-plane momenta for the $m$=$-$1,0,$+$1 grating orders that allow coupling to the SPhPs at intersections with the SPhP dispersion (green). d) Zoom into the counter-propagating ($k_\parallel<0$) SPhP branch, highlighting the steep dispersion expected from the $m$=$-$1 grating order.}
  \label{fig1}
\end{figure*}

The experimental scheme is outlined in Fig.~\ref{fig1}a. The IR laser pulses generated by the IR free-electron laser (FEL) installed at the Fritz Haber Institute impinge on the sample, and the reflected IR intensity as well as the SHG are detected. The FEL provides high-power, narrow-band (bandwidth of $\sim 0.3$\%), and wavelength-tunable IR light, and is described in detail elsewhere.\cite{Schollkopf2015} 

The IR laser beam was mildly focused onto the sample (spot size $\approx$ 600~$\mu$m) with angle of incidence $\alpha$, polarized in the \emph{x-z} plane (P-polarized) with the lines of the grating along the \emph{y}-direction, see Fig.~\ref{fig1}a. The reflected fundamental beam as well as the SHG beam emitted in specular reflection are each detected using home-built pyroelectric and mercury cadmium telluride detectors (IR associates), respectively, after spectral separation using IR short-pass filters. SHG and linear reflectance spectra are acquired by spectral scanning of the FEL wavenumber from 770 to \wavenumber{1020} using the motorized undulator gap.

The gold line grating samples were produced on a c-cut semi-insulating 6H-SiC substrate using 3D two-photon laser lithography printing\cite{Hohmann2015}. In this technique, a wavelength-dependent photoresist is uniformly applied to the substrate, and the grating is patterned by scanning the focus of an illuminating laser across. After a developing process, a gold layer of $h$=$200$~nm was deposited with an electron beam evaporator and a final lift-off was done.. Two separate samples where made, one with grating line width $d$=$2.4~\mu$m and one with $d$=$4.8~\mu$m, however, both with the same period $p$=$7.2~\mu$m. An optical microscope image of the $d$=$4.8~\mu$m sample is shown in Fig.~\ref{fig1}b.

Propagating polaritons can be excited in these structures within the Reststrahlen band of SiC between the transversal optical (TO) and longitudinal optical (LO) phonon frequencies where the dielectric permittivity is negative.\cite{Caldwell2013a,Chen2014,Razdolski2016} The dispersion relation $k(\omega)$ of the extraordinary surface phonon polaritons at the $z$-cut uni-axial crystal surface is given as:\cite{Otto1968}
\begin{equation}
    k_{SPhP} = \frac{\omega}{c_0}\sqrt{\frac{\epsilon_\perp\epsilon_\parallel - \epsilon_\parallel}{\epsilon_\perp\epsilon_\parallel -1}},
    \label{eq:SPhP}
\end{equation}
where $k$ and $\omega$ are the in-plane momentum and the wavenumber, respectively, and $\epsilon_\perp(\epsilon_\parallel)$ is the in-plane (out-of-plane) dielectric function of 6H-SiC.\cite{Passler2017} The dispersion of forward and backward propagating surface phonon polaritons at the 6H-SiC surface are plotted as green solid lines in Fig.~\ref{fig1}c and d.  

Momentum transfer from a line grating, in the simplest description, can be modelled by adding the grating momentum $G_m=\frac{m}{p}$ with the grating period $p$ and the diffraction order $m$ to the in-plane momentum component of the incoming light $k_x(\alpha)$. This leads to a total in-plane momentum $k(\alpha,m) = k_x(\alpha) + G_m$, as illustrated in Fig.~\ref{fig1}c for $m$=$-$1,0,$+$1 for a range of incidence angles $\alpha$ of the incoming light. Surface polaritons can be excited if:
\begin{equation}
    k_{SPhP} = k_\parallel(\alpha,m) = k_x(\alpha) + \frac{m}{p}.
    \label{eq:kick}
\end{equation}
Clearly, no coupling is possible for $m$=0 where $k_\parallel$ remains within the light cone. For the period $p=7.2~\mu$m chosen in our work, the $m$=$+$1 order provides access to large momentum modes with small dispersion, while a steep dispersion is expected for the $m$=$-$1, counter-propagating diffraction order. The latter is illustrated in Fig.~\ref{fig1}d, where a steep dependence of the polariton resonance wavenumber on incidence angle is predicted. 

\begin{figure}[bth!]
\includegraphics[width=\columnwidth]{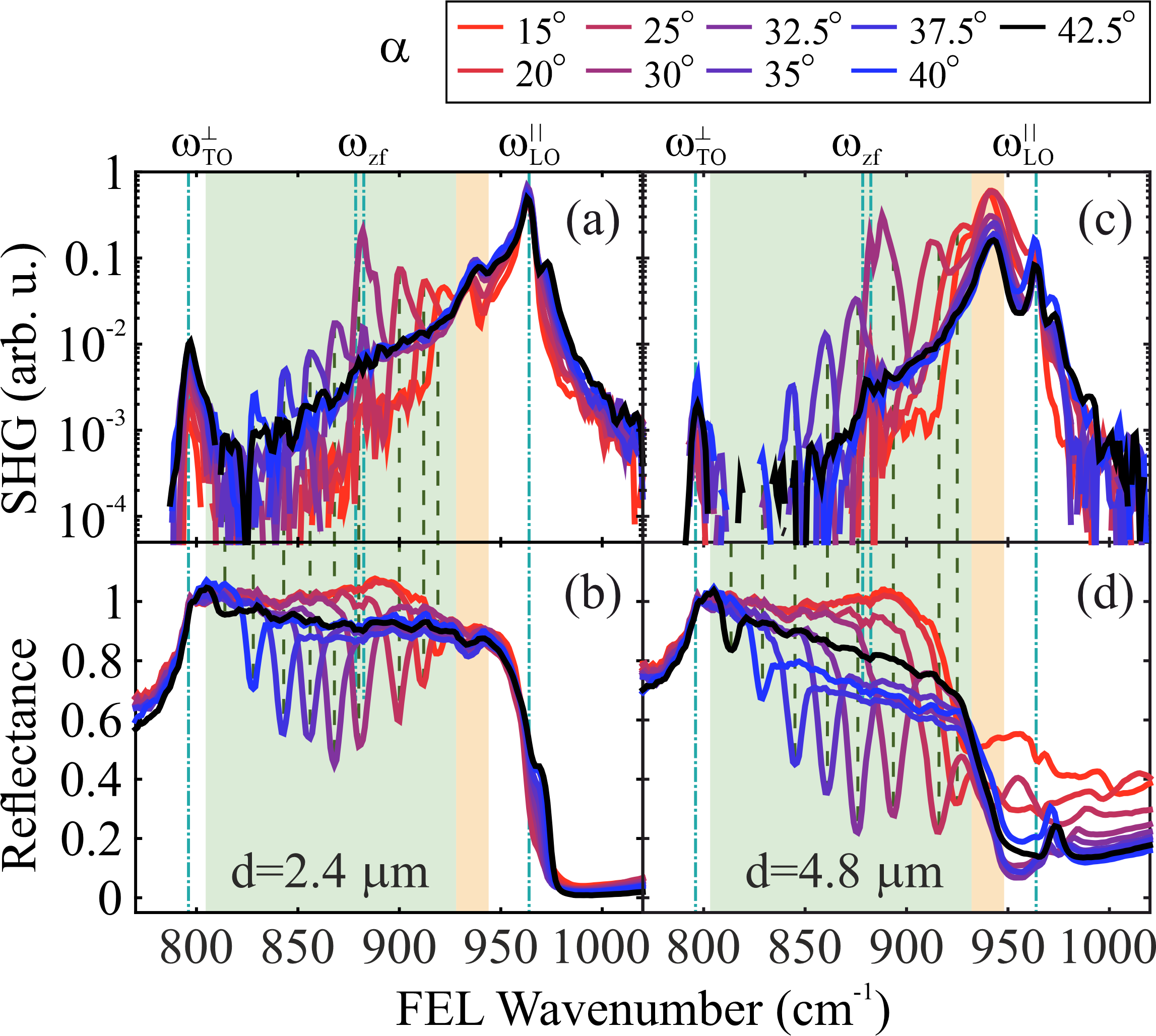}
  \caption{\textbf{Experimental SHG and IR reflectance spectra.} SHG intensity (a,c) and reflectance (b,d) at various incidence angles for $d$=$2.4$~$\mu$m (a,b) and $d$=$4.8$~$\mu$m (c,d). Dips in the reflectance with associated peaks in the SHG strongly dispersing with incidence angle (red-to-blue) are assigned to resonant excitation of surface polaritons via the $m$=$-$1 grating order (light green shaded area). Non-dispersive features include 6H-SiC substrate resonances at $\omega_{TO}^\perp$ and $\omega_{LO}^\parallel$, and the zone-folded modes $\omega_{zf}$ previously observed in SHG spectroscopy of SiC.\cite{Paarmann2016,Razdolski2016} Additionally, a weakly dispersive feature at $\omega \approx$\wavenumber{940} is tentatively assigned to surface polariton excitation via the $m$=$+$1 grating order (light orange shaded area). 
  Please note the logarithmic scale in a) and c).} 
  \label{fig2}
\end{figure}

Experimental SHG and reflectance spectra at various incidence angles for both samples are shown in Fig.~\ref{fig2}. The SHG spectra feature several resonances: the peaks at the LO phonon wavenumber $\omega_{LO}^\parallel$ and the TO phonon wavenumber $\omega_{TO}^\perp$ are typical for SHG spectroscopy of SiC and arise due to resonances in the Fresnel transmission and the nonlinear susceptibility $\chi^{(2)}$, respectively.\cite{Paarmann2015,Paarmann2016,Passler2017} These features also mark the edges of the Reststrahlen band of SiC observed in the reflectance spectra, Fig.~\ref{fig2}b. Notably, these features are almost unaffected by the incidence angle. In the reflectance spectra for $d$=$4.8$~$\mu$m, Fig.~\ref{fig2}d, we find the upper Reststrahlen band edge significantly modified due to the large total area of gold on the sample surface.

The strongly dispersive peaks in the SHG spectra and dips in the reflectance spectra between \wavenumber{810} and \wavenumber{920} indicate resonant excitation of counter-propagating surface polaritons via the $m$=$-$1 grating order, in qualitative agreement with the prediction in Fig.~\ref{fig1}d. Hereby, the amplitude of the SHG enhancement correlates with the magnitude of the reflectance dip, reaching enhancement values of about 10. An additional tenfold SHG  enhancement is observed for the polariton resonance in spectral proximity of the zone-folded weak modes of 6H-SiC at $\omega_{zf}=$\wavenumber{881,886},\cite{Bluet1999,Paarmann2016} similar to what was observed for localized polaritons in subdiffractional nanostructures.\cite{Razdolski2016} 

Finally, a weakly dispersive SHG resonance is observed at \wavenumber{940}. For $d$=$2.4$~$\mu$m in Fig.~\ref{fig2}a and b, the weak SHG peak correlates with a small dip in the reflectance, suggesting (relatively inefficient) excitation of forward-propagating polaritons via the $m$=$+$1 grating order. For $d$=$4.8$~$\mu$m, the corresponding SHG peak is much more pronounced. In the reflectance spectra for this sample, Fig.~\ref{fig2}d, a strong, broad bleach is observed which, however, cannot be clearly assigned  to a polariton resonance. 

\begin{figure}[ht!]
\includegraphics[width=\columnwidth]{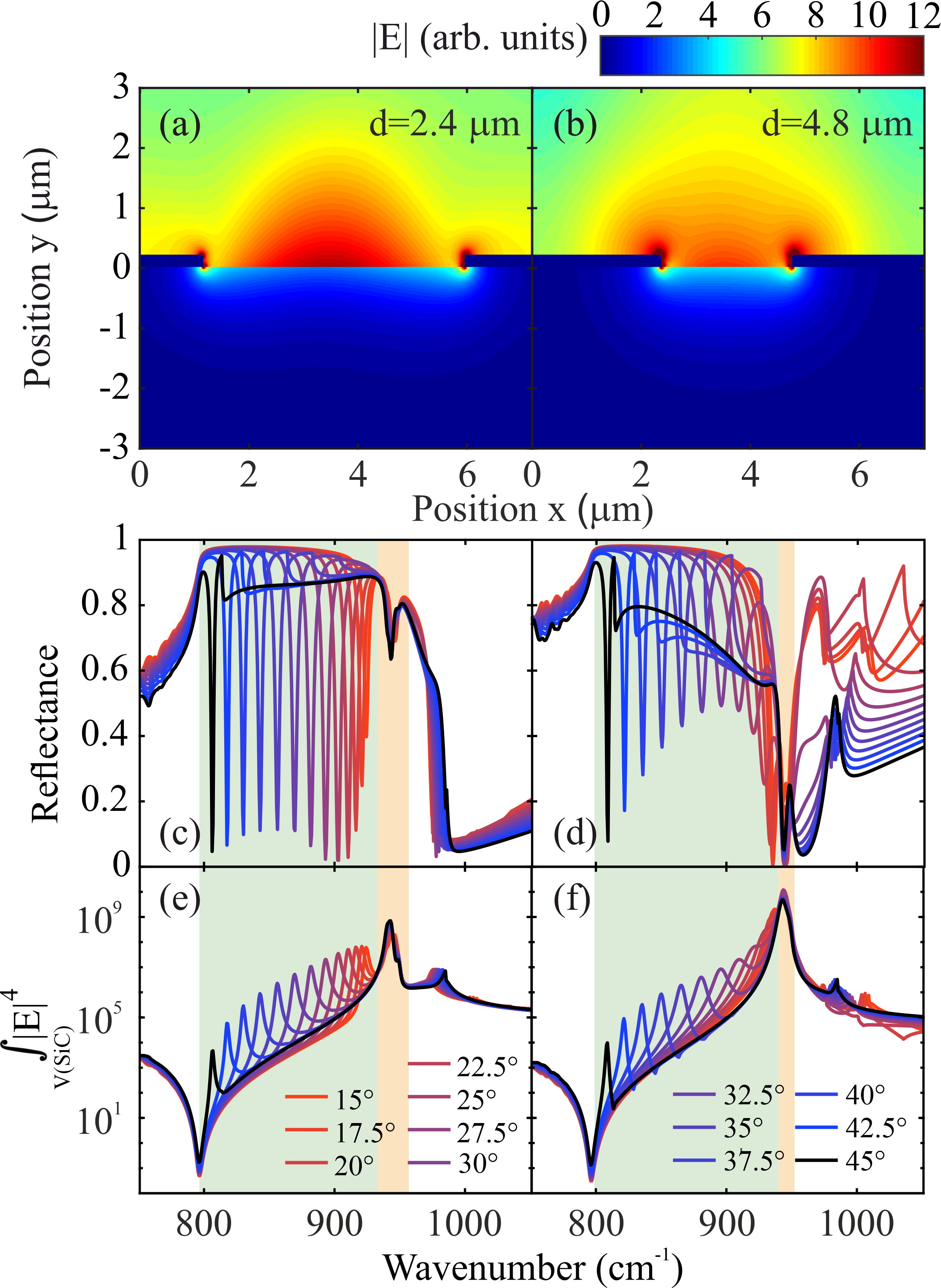}
  \caption{\textbf{Simulations of grating coupling to surface polaritons.} Electric field distributions for a) $d$=$2.4$~$\mu$m and b) $d$=$4.8$~$\mu$m at the respective $m$=$-$1 polariton resonance for $\alpha = 30^\circ$, exhibiting both homogeneous as well as strongly localized contributions of the polaritonic field enhancement. c,d)  Calculated reflectance spectra for various incidence angles (see legend). e,f) Spectra of the fourth power of the field enhancement driving SHG integrated over the SiC volume, $\int_{V(SiC)}|E|^4$. Light green and light orange shaded areas mark the spectral ranges for the backward-propagating $m$=$-$1 and forward-propagating $m$=$+$1 grating orders, respectively.} 
  \label{fig3}
\end{figure}

To corroborate these experimental findings, we performed finite-element simulations of our structures using Comsol Multiphysics.\cite{Comsol2021} Frequency-domain calculations with periodic Floquet boundary conditions including first order diffraction yielded the 
frequency-dependent electric field distributions as well as the reflectance spectra for various incidence angles, see Fig.~\ref{fig3}a,b and c,d, respectively, for the two grating geometries studied experimentally. Hereby, we used the isotropic dielectric function of SiC\cite{Paarmann2016} and the Drude model for gold.   

For $d$=$2.4$~$\mu$m, the deep dips in the reflectance spectra, Fig.~\ref{fig3}c, dispersing with incident angle suggest near-optimal sample design parameters for excitation of the counter-propagating polaritons via the $m$=$-$1 grating order, in excellent qualitative agreement with the experiment. Similarly, the shallow dip with small dispersion at \wavenumber{940}  suggests excitation of forward-propagating polaritons via the $m$=$+$1 grating order, though with less efficiency just like in the experiment. We also note the step-like drop of the reflectance arising at wavenumbers slightly above the $m$=$-$1 polariton resonance which corresponds to onset of first order diffraction of the sub-wavelength period grating, \emph{i.e.}, the Wood's anomaly.\cite{Wood1902} We did not observe clear signatures of the Wood's anomaly in the experiment, and will thus focus on the polariton resonances in the following.   

The electric field distributions for resonant counter-propagating $m$=$-$1 polariton excitation under incidence angle $\alpha=30^\circ$ at $\omega$=\wavenumber{884} and $\omega$=\wavenumber{894} is plotted in Fig.~\ref{fig3}a and b, respectively, for $d$=$2.4$~$\mu$m and $d$=$4.8$~$\mu$m. Very similar field distributions are found at the counter-propagating polariton resonance at different incidence angles and at the $m$=$+$1 resonances (both not shown). Fig.~\ref{fig3}a and b suggest two major components of the polaritonic field enhancement;  one being relatively homogeneous along both sides of the SiC/air interface as expected for a propagating polariton, and another being largely localized in a much smaller volume at the edges of the gold grating. These latter hot spots already suggest that the gold grating does not merely serve as a momentum transducer for SPhP excitation, but instead actively contributes to the formation of the polariton mode.

For $d$=$4.8$~$\mu$m, the reflectance spectra in Fig.~\ref{fig3}d behave quite similar to $d$=$2.4$~$\mu$m in terms of the $m$=$-$1 polariton resonance. However, a clearly different behavior is observed for $\omega>\wavenumber{940}$, similar to what was found in the experiment. First, the reflectance drops sharply at $\omega=$\wavenumber{940} around where the $m$=$+$1 polariton resonance is predicted. Second, significantly higher reflectance values are found above the SiC Reststrahlen band, indicative of an optical response being strongly modified by the presence of the gold grating in this range. Specifically interesting is the rapid flip from peak to dip from $\alpha<30^\circ$ to $\alpha>30^\circ$ at the upper edge of the SiC Reststrahlen band, $\omega=$\wavenumber{965}, yet again in excellent qualitative agreement with the experimental reflectance. 

\begin{figure}[ht!]
\includegraphics[width=\columnwidth]{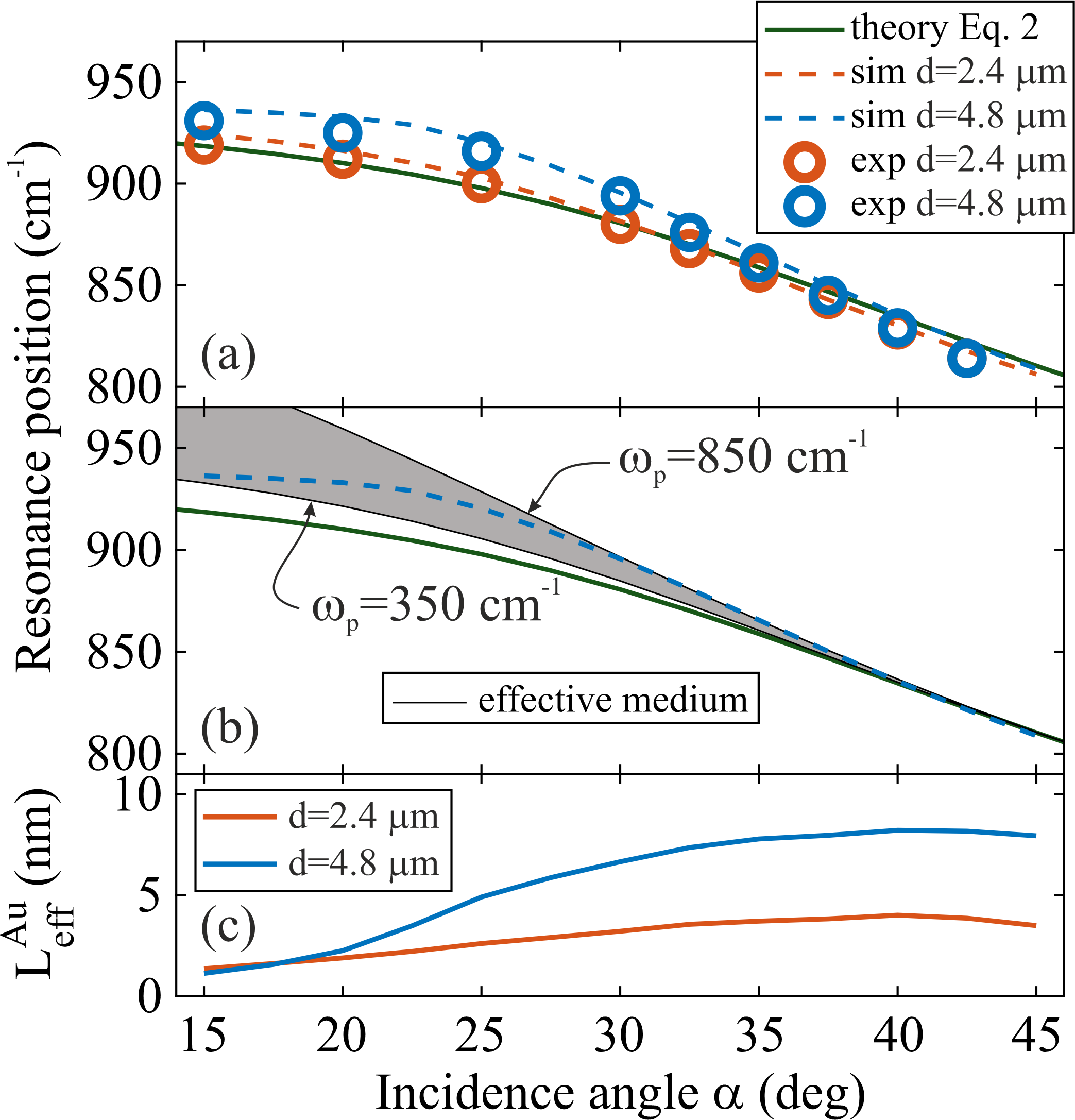}
  \caption{\textbf{Hybridization of grating-coupled surface polaritons} a) Experimental (circles) and simulated (dashed lines) dispersion of $m$=$-$1 grating coupled polaritons extracted from the dip positions in the reflectance spectra for $d=$2.4~$\mu$m (blue) and $d=$4.8~$\mu$m (red). The data for $d=$2.4~$\mu$m agrees well with the theoretical dispersion, Eq.~\ref{eq:kick}, while a systematic blue shift is observed for $d=$4.8~$\mu$m. b) Comparison of the simulated $d=$4.8~$\mu$m dispersion with effective medium modelling using Eq.~\ref{eq:EM} with $\omega_p$=\wavenumber{350} (lower grey edge) and $\omega_p$=\wavenumber{850} (upper grey edge), matching the simulated $m$=$-$1 dispersion at $\alpha$=15$^\circ$ and for $\alpha>$25$^\circ$, respectively. This can be interpreted as a change in the effective plasma frequency $\omega_p$ required to reproduce the simulation along the dispersion. This behavior correlates well with the effective mode length\cite{Maier2006} $L_{eff }^{Au}$ within the gold shown in c), such that the number of charge carriers participating in the mode reduces for larger momenta (smaller incidence angles). }
  \label{fig4}
\end{figure}

With the overall good agreement of the simulated reflectance spectra with the experiment, we can confidently use the simulation results to further analyze the SHG spectra. A microscopic description of SHG\cite{Razdolski2016} in this structure is beyond the scope of this work. Instead, we here estimate the contribution of the polaritonic field enhancement to the SHG yield by integrating the fourth power of the electric field magnitude across the SiC volume, $\int_{V(\mathrm{SiC})}|E|^4$. By doing so, we neglect any phase relation between the local sources of nonlinear signal and the resulting impact on the far-field emission, as well as the tensorial nature of the nonlinear susceptibility.  

The spectral dependence of $\int_{V(\mathrm{SiC})}|E|^4$ is plotted in Fig.~\ref{fig3}e,f. These simulated data resemble the experimental SHG spectra, see Fig.~\ref{fig2}a and c, with remarkable precision, not only in terms of the spectral position of the resonances, but also in terms of the relative amplitudes between peaks and background. Even the pronounced enhancement of the SHG yield at the $m$=$+$1 polariton resonance between $d$=$2.4$ and $4.8$~$\mu$m is almost perfectly reproduced in the simulations. Notably, the SHG resonance at the TO phonon frequency $\omega_{TO,\perp} =$\wavenumber{797} arises due to the resonance in the nonlinear susceptibility of SiC\cite{Paarmann2015,Paarmann2016} and would, thus, not be expected to appear in the field enhancement where instead a deep dip reports on the Fresnel anti-resonance\cite{Paarmann2016} at the TO phonon. 

Notably, we observe a blue-shift of the dispersive $m$=$-$1 polariton resonance between the $d$=2.4 and 4.8~$\mu$m geometries, both in the experiment and in the simulations. This is shown in Fig.~\ref{fig4}a where we plot the polariton resonance positions we extracted from the dips in the reflectance spectra. For each grating line width, experiment and simulations agree very well. For reference, we also plot the theoretical dispersion, Eq.~\ref{eq:kick}. 

For $d$=$2.4$~$\mu$m, the experimental and simulated dispersion agree almost perfectly with the theoretical prediction, suggesting that here the polariton mode is dominated by the SiC properties and the influence of the gold is rather weak. In contrast, for $d$=$4.8$~$\mu$m the dispersion is stretched to higher wavenumbers,\emph{i.e.}, the blue-shift increases for larger $|k_\parallel|$, reaching values of $>$\wavenumber{15} at $\alpha=15^\circ$. Clearly, the simple model of momentum transfer by the grating coupler, Eq.~\ref{eq:kick}, fails to accurately describe the dispersion, which we ascribe to mode hybridization between the SPhP and the local response of the free charge carriers at the gold grating etches,\emph{i.e.}, localized plasmons.    

A similar behavior of plasmon-phonon polariton hybridization has been observed recently in studies of surface polaritons in heavily doped semiconductors\cite{Shalygin2019,Janonis2020}, van der Waals heterostructures,\cite{Dai2015,Woessner2015,Huber2016} and metallic nanostructures placed on polar crystals slabs\cite{Huck2016,Pons-Valencia2019}, with varying degrees of coupling strength. For the case of doped semiconductors, the plasmon-phonon coupling can be described analytically.\cite{Shalygin2019} Inclusion of a Drude term in the SiC dielectric function describing the free charge carrier response results in an additional pole in the dielectric function, while the pole at the LO phonon frequency is blue-shifted. In consequence, the Reststrahlen-band broadens and the polariton dispersion is stretched.\cite{Shalygin2019,Janonis2020} 

We can use this simple model for an effective medium description of the Au grating-induced plasmonic contribution to the polariton dispersion, by adding a Drude term to the dielectric function of SiC:
\begin{equation}
    \epsilon_{\perp(\parallel)}^{\mathrm{eff}}(\omega) = \epsilon_{\perp(\parallel)}^{SiC}(\omega) - \epsilon_{\infty,\parallel(\perp)}^{SiC}\left(\frac{\omega_p^2}{\omega^2+i\gamma\omega}\right),
    \label{eq:EM}
\end{equation}
where $\epsilon_{\perp(\parallel)}^{SiC}(\omega)$ and $\epsilon_{\infty,\parallel(\perp)}^{SiC}$ are the in-plane (out-of-plane) intrinsic dielectric function and high-frequency permittivity of SiC, $\omega_p$ the plasma frequency and $\gamma$ the plasma damping. The resulting polariton dispersion is plotted in Fig.~\ref{fig4}b for  $\omega_p$=\wavenumber{350} and $\omega_p$=\wavenumber{850} with $\gamma$=$0.1\omega_p$. While the former matches the simulated dispersion for large momenta (small $\alpha$), the latter better describes the initial slope of the dispersion (larger $\alpha$). Thus, the effective medium  description using a single plasma frequency cannot sufficiently describe the plasmon-phonon polariton hybridization.

One possibility to explain this discrepancy could be the different magnitude of the plasmon contribution to the hybrid mode along the dispersion. To test this hypothesis, we evaluated the effective mode volume\cite{Maier2006} within the Au from the finite-element simulations Fig.~\ref{fig3}. In this picture, the different spatial extent of mode penetration into the gold results in a different effective number of charge carriers actively participating in the mode, leading to a different effective plasma frequency in Equ.~\ref{eq:EM}. Indeed, the extracted effective mode lengths $L_{\mathrm{eff}}^{\mathrm{Au}}$ shown in Fig.~\ref{fig4}c agree very well with our observations: For $d$=$2.4$~$\mu$m we find only a mild change of $L_{\mathrm{eff}}^{\mathrm{Au}}$ along the dispersion. For $d$=$4.8$~$\mu$m, however, $L_{\mathrm{eff}}^{\mathrm{Au}}$ increases significantly with $\alpha$, i.e., the mode penetrates deeper and deeper into the gold as the polariton momentum is reduced towards the light line.  

We note that frequency shifts of the polariton resonances can additionally occur due to radiative coupling, which typically results in a broadening and simultaneous red-shift of the polariton resonances.\cite{Passler2017} Fig.~\ref{fig3}d indeed suggests increasing radiative coupling with decreasing $\alpha$ (increasing $|k_\parallel|$) for the $m$=-1 resonance, where the reflectance dip amplitudes decrease while the dip widths increase. Such a behavior is in fact identical to the critical coupling effect in the Otto-type prism coupling configuration.\cite{Passler2017} These red-shifts cannot, however, explain the blue-shift of the hybrid polariton resonance observed in Fig.~\ref{fig4}a. Careful design of the grating structure might allow to simultaneously optimize, e.g., plasmon hybridization and radiative coupling across the dispersion. For instance, a thickness gradient in the gold film across each grating stripe would provide an additional design parameter to independently control both quantities.  

Overall, the polaritonic SHG enhancement we find here using the grating coupling configuration is much less than what we previously observed in subdiffractional nanostructures\cite{Razdolski2016} and using Otto-type prism coupling\cite{Passler2017}, where in both cases the polariton resonances dominated the SHG spectra, and where we generally observed larger SHG signal amplitudes. In fact, we expect that formation of hot spots would boost the SHG in contrast to homogeneous polariton field distributions typical for prism coupling\cite{Passler2017} because of the fourth order electric field scaling of the nonlinear signal, in contrast to our observations. In principle, interference between the hot spot emissions as well as shadowing by the gold could reduce the far-field coupling of the SHG. However, the excellent agreement of the simulations in Fig.~\ref{fig3}e,f with the experimental SHG spectra suggests that the reduced yield occurs already in the generation process. Further experiments and simulations including SHG far-field coupling are needed to clarify the lack of efficiency of the SHG in the grating geometry.

In summary, we demonstrated infrared SHG from grating-coupled polaritons using gold gratings on 6H-SiC. We show that the polariton resonance can be widely tuned using the incidence angle in the counter-propagating $m$=$-$1 grating order. Unfortunately, we find the yield of the polaritonic enhancement of the SHG to be rather small, compared to previous work using prism-coupled or localized phonon polaritons. Further, we find that the polariton dispersion can be significantly modified using different width-to-gap ratios of the grating, which we ascribe to hybridization of the propagating phonon polariton modes with localized plasmon resonances in the gold grating. We qualitatively describe the modification of the hybrid mode dispersion with an effective medium description which also accounts for the variation of the effective mode volume within the gold along the dispersion. We propose that this platform is ideal to engineer the hybrid mode dispersion. Conveniently, gold gratings can be employed for any polar crystal surface, providing a versatile tool for probing hybrid polaritons in a wide range of polariton materials. 

\begin{acknowledgements}

The authors wish to thank S.~Gewinner and W.~Sch\"ollkopf for operating the FEL. 

\end{acknowledgements}

\section*{Data Availability Statement}

The data that support the findings of this study are openly available in Zenodo at http://doi.org/10.5281/zenodo.6880979.

\nocite{*}

\bibliography{kohlmann_SHG_grating}

\end{document}